# Soliton spiraling in optically-induced rotating Bessel lattices

Yaroslav V. Kartashov, Victor A. Vysloukh, Lluis Torner

*ICFO-Institut de Ciencies Fotoniques and Department of Signal Theory and Communications, Universitat Politecnica de Catalunya, 08034 Barcelona, Spain*

We address soliton spiraling in optical lattices induced by multiple coherent Bessel beams and show that the dynamical nature of such lattices make them able to drag different soliton structures, setting them into rotation. The rotation rate can be controlled by varying the topological charges of lattice-inducing Bessel beams.

Light propagation in media with modulated refractive index exhibits a variety of unique phenomena. When the modulation forms an array of evanescently coupled waveguides, discrete solitons appear as excitations of a few guides.[1] Discrete solitons can move across the array, a phenomenon accompanied by radiation but with important all-optical routing applications. Tuning the depth of the refractive-index modulation is a powerful tool to control the soliton mobility.[2] Such tunable lattices can be generated optically in photorefractive media.[3-6]

To date, soliton mobility has been analyzed in lattices that keep their shape invariant along the longitudinal direction. In this Letter we explore, for the first time to our knowledge, optically induced lattices whose shape evolves dynamically. In particular, we address a phenomenon of soliton spiraling in a lattice that drags solitons into rotary motion. Such lattices can be generated by the interference of several Bessel beams.[7] The resulting refractive-index modulation produces guiding structures that resemble optical fibers with a varying shape. However, the possibility to reconfigure and optically tune properties of such structures is a central point of the concept described here. We discuss the impact of lattice strength on soliton motion as well as the possibility of controlling the soliton rotation rate by changing topological charges of lattice-creating beams.

We consider light propagation along the $z$ axis of focusing cubic medium with transverse modulation of refractive index, described by the nonlinear Schrödinger equation for dimensionless complex field amplitude $q$:

$$i\frac{\partial q}{\partial \xi} = -\frac{1}{2}\left(\frac{\partial^2 q}{\partial \eta^2} + \frac{\partial^2 q}{\partial \zeta^2}\right) - q|q|^2 - pR(\eta,\zeta,\xi)q. \qquad (1)$$

Here the longitudinal $\xi$ and transverse $\eta,\zeta$ coordinates are scaled to the diffraction length and input beam width, respectively; the parameter $p$ is proportional to the refractive index modulation depth; the function $R(\eta,\zeta,\xi)$ describes the profile of the lattice. Among the conserved quantities of Eq. (1) is the energy flow $U = \int_{-\infty}^{\infty}\int_{-\infty}^{\infty}|q|^2\,d\eta d\zeta$.

We address refractive-index modulations induced optically with interfering Bessel beams $J_k[(2b_k)^{1/2}r]\exp(ik\phi - ib_k\xi)$ with different topological charges $k$. Here $r^2 = \eta^2 + \zeta^2$ is the radius, $\phi$ is the azimuth angle, the parameter $b_k$ defines the transverse scale of the $k$-th beam. Good approximations to Bessel beams can be created experimentally by a variety of techniques.[8,9] We assume that the lattice shape features the intensity of the interference pattern $R(\eta,\zeta,\xi) \sim \sum_k J_k^2 + 2\sum_n\sum_{m>n} J_n J_m \cos[(n-m)\phi + (b_m - b_n)\xi]$ of several beams. In the simplest case of two Bessel beams with $b_n, b_m$ chosen in such way that both functions $J_n[(2b_n)^{1/2}r]$ and $J_m[(2b_m)^{1/2}r]$ acquire their maximal values at $r=1$, one gets a twisted lattice with a rotation rate defined by $b_m - b_n$, and an azimuthal symmetry defined by value of $n-m$. Except for the rotation, the cross-section of such lattice is invariant along $\xi$. There are two characteristic longitudinal scales for the twisted lattices: the total rotation period $T_{\rm rot} = 2\pi(m-n)/(b_m - b_n)$ and the distance $T = 2\pi/(b_m - b_n)$, where the lattice shape replicates itself. For the simplest lattice with $m-n=1$ one has $T_{\rm rot} = T$. Even though the induced refractive index modulation decreases as $r \to \infty$ we use the term *optical lattice* to stress a possibility to reconfigure it all-optically. Notice that the infinite extent of harmonic lattices is only important in linear propagation, when many lattice sites are occupied, in contrast to the nonlinear case addressed here.

To study the soliton dragging we fixed the charge $n=1$ and varied $m$. For $m=2$ the twisted lattice has one clearly pronounced guiding channel (Fig. 1(a)); in that case we set the input field distribution in the form of a single soliton supported by the lattice at $\xi = 0$. Such beam would propagate undistorted in the non-rotating lattice. The soliton beam is characterized by its energy flow $U$. Narrow solitons cease to feel the lattice and their energy flow approaches that of Townes solitons: $U_{\max} = 5.85$. In the twisted dynamical lattices such beams are no longer stationary, but for deep enough lattices they can be set into steady spiral motion with a period dictated by the period of the lattice rotation (Fig. 1). Such soliton motion is conceptually different from previous works on soliton steering and spiraling,[10,11] since here the otherwise immobile solitons are dragged by the dynamical lattice. We found that solitons can follow the lattice for huge distances, exceeding any feasible crystal length and undergo thousands of rotations. The location of the soliton peak amplitude exhibits small oscillations inside a rotating guiding channel of the lattice (Fig. 1). In the case of a lattice-creating beam with charge $m = 3$ (Fig. 2), two pronounced guiding channels can be used to drag more complicated soliton structures such as dipole-mode solitons. We have found the input profile of dipole-mode soliton as exact solution of Eq. (1) at $\xi = 0$. The limiting energy flow for existence of dipole-mode solitons is given by $2U_{\max}$ and close to this value the dipole-mode solitons transform into two narrow and almost non-interacting solitons. Dipole-mode soliton are also set into rotary motion by the lattice (Fig. 2). Besides dipole-mode soliton the lattice shown in Fig. 2 can drag single soliton located in either of two guiding sites.

    To show that nonlinearity is essential for steady soliton spiraling we compared propagation of identical beams in the lattice shown in Fig. 1 with and without nonlinearity (Fig. 3(a), (b)). In the linear regime radiation losses grow dramatically, the beam broadens and is redistributed between secondary lattice rings, thus quickly spreading after several lattice rotations. In the nonlinear regime the radiation is orders of magnitude smaller. To quantify the radiation rate we calculated the energy flow $U_\xi$ concentrated in a close proximity of main guiding lattice sites (within the ring shown by dashed lines in Figs. 1(a), 2(a)) for different propagation distances (Fig. 4(a)). Radiation rate is larger during several initial rotations but it drastically reduces with increase of $\xi$, and beyond $\xi = 150$ the total radiative losses over a hundred diffraction lengths (around 50 rotations) amount to about one percent. Similar result was obtained for dipole-mode solitons dragged by the lattice shown in Fig. 2.

We also found that the radiation rate decreases with growth of the lattice depth and the input soliton energy flow. To quantify this effect we set the criterion that soliton is captured by the lattice when radiative losses after one rotation (for dipole-mode soliton after one self-replication period) amount to less than 5%. Since losses decrease monotonically with increase of the input energy flow, this enables to define the critical energy level $U_{\rm cr}$. The critical energy decreases with growth of the lattice depth (Figs. 4(b) and 4(c)). This implies that deep enough lattices become totally trapping and can drag even low-energy beams. However, here we only interested in parameter range where nonlinearity really affects dragging properties of the lattice as it is shown in Figs. 3(a) and 3(b). Note that typically at $U \approx U_{\rm cr}$ nonlinearity strength is comparable with that of the lattice. With decrease of lattice depth $U_{\rm cr}$ approaches $U_{\rm max}$ for single solitons and $2U_{\rm max}$ for dipole-mode solitons. This implies that below certain critical value of $p$, the lattice may be unable to trap and drag solitons. This critical value ($\sim 18$ for rapidly rotating lattices considered here) is reduced considerably for lattices with lower rotation rates formed by broader Bessel beams. The maximal rotation rate of twisted lattice is dictated by the minimal achievable size of the core of lattice-creating beam, which is of the order of wavelength. Lattices created with such beams are expected to perform tens of rotations per millimeter.

A key result is that by changing the topological charge of one of lattice-creating beams, it is possible to control the rotation rate of the lattice, and, hence, the output soliton position at a given propagation length (Fig. 4(d)). Besides changing the rotation rate, an increase of the topological charge $m$ results in more complex refractive index distributions (Fig. 3(c),(d)). Such lattices can trap and drag soliton complexes composed from several bright spots as well as single beams located in either of guiding channels of the structure.

In conclusion, the results reported here introduce an important new concept for soliton control. The new scheme is based on the dragging of different soliton structures that is caused by dynamically varying optical lattices made by multiple interfering Bessel beams. The approach has several control parameters, including the depth of the lattice, and the topological charge of the lattice-creating beams.

# Figure captions

Figure 1. Soliton propagation in the lattice with $m=2$. Left and right columns show refractive index and field distributions at various distances. White rings in the right column display a trajectory of the lattice maximum and arrows show rotation direction. Input energy flow $U_{\text{in}}=1.5$ and $p=32$.

Figure 2. The same as in Fig. 1 but for dipole-mode soliton in the lattice with $m=3$ at $U_{\text{in}}=1.5$, $p=50$.

Figure 3. Output field distributions after two rotations of the lattice with $m=2$ in nonlinear (a) and linear (b) medium at $U_{\text{in}}=4$, $p=25$. Lattices corresponding to $m=4$ (c) and $m=5$ (d).

Figure 4. (a) Energy flow of single soliton captured by the lattice with $m=2$ versus propagation distance at $U_{\text{in}}=1.5$, $p=32$. Critical energy versus lattice depth for single soliton captured by the lattice with $m=2$ (b) and dipole-mode soliton captured by the lattice with $m=3$ (c). (d) Rotation $T_{\text{rot}}$ and self-replication $T$ periods versus charge $m$.

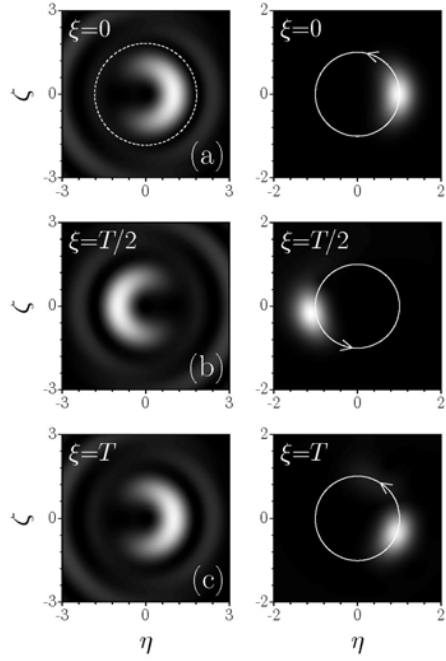

Figure 1. Soliton propagation in the lattice with $m=2$. Left and right columns show refractive index and field distributions at various distances. White rings in the right column display a trajectory of the lattice maximum and arrows show rotation direction. Input energy flow $U_{\rm in}=1.5$ and $p=32$.

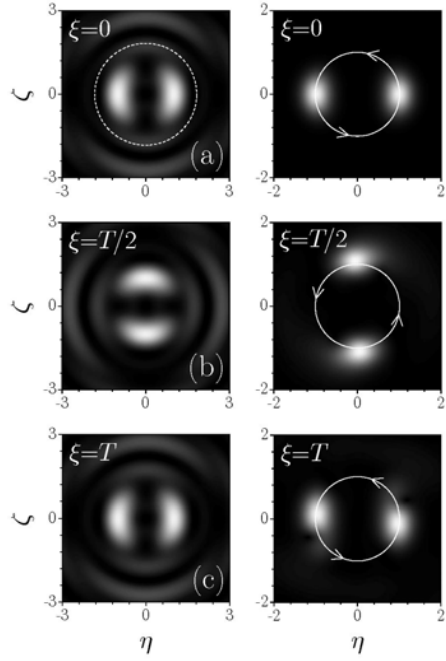

Figure 2. The same as in Fig. 1 but for dipole-mode soliton in the lattice with $m=3$ at $U_{\text{in}}=1.5$, $p=50$.

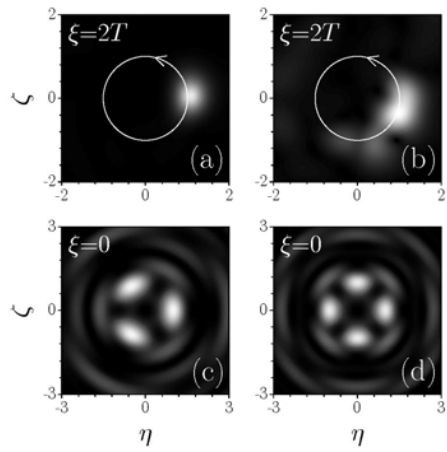

Figure 3. Output field distributions after two rotations of the lattice with $m = 2$ in nonlinear (a) and linear (b) medium at $U_{\text{in}} = 4$, $p = 25$. Lattices corresponding to $m = 4$ (c) and $m = 5$ (d).

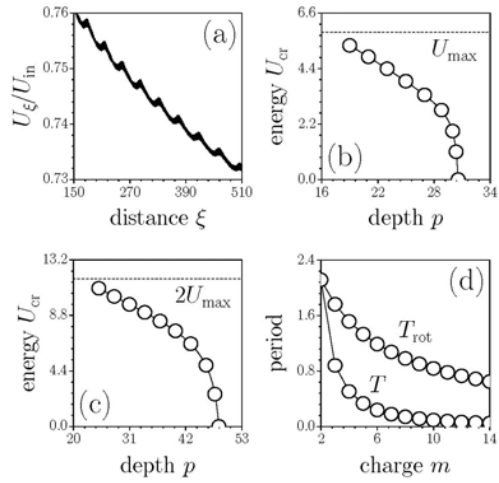

Figure 4. (a) Energy flow of single soliton captured by the lattice with $m=2$ versus propagation distance at $U_{in}=1.5$, $p=32$. Critical energy versus lattice depth for single soliton captured by the lattice with $m=2$ (b) and dipole-mode soliton captured by the lattice with $m=3$ (c). (d) Rotation $T_{rot}$ and self-replication $T$ periods versus charge $m$.